\documentclass[floatfix,aps,twocolumn,showpacs,superscriptaddress,10pt,longbibliography]{revtex4-2}
\usepackage{amsmath,amsfonts,amssymb}
\usepackage{graphicx} 
\usepackage{epstopdf}
\usepackage{bm}
\usepackage{hyperref}

\usepackage[utf8]{inputenc}
\usepackage[T1]{fontenc}
\usepackage[english]{babel}

\IfFileExists{newtxtext.sty}
 {\usepackage{newtxtext,newtxmath}}
 {\IfFileExists{stix.sty}
  {\usepackage{stix}}
  {\IfFileExists{mathptmx.sty}
  {\usepackage{mathptmx}}{} } }

\newcommand{\rrangle}{\rangle\!\rangle}
\newcommand{\llangle}{\langle\!\langle}

\begin{document}
\title{Unraveling Open Quantum Dynamics with Time-Dependent Variational Monte Carlo}

\author{Christian Apostoli}
\thanks{These authors contributed equally to this work.}
\affiliation{Dipartimento di Fisica ``Aldo Pontremoli'', Universit\`a degli Studi di Milano, via Celoria 16, I-20133 Milano, Italy}
\author{Jacopo D'Alberto}
\thanks{These authors contributed equally to this work.}
\affiliation{Dipartimento di Fisica ``Aldo Pontremoli'', Universit\`a degli Studi di Milano, via Celoria 16, I-20133 Milano, Italy}
\author{Marco G. Genoni}
\affiliation{Dipartimento di Fisica ``Aldo Pontremoli'', Universit\`a degli Studi di Milano, via Celoria 16, I-20133 Milano, Italy}
\affiliation{INFN, Sezione di Milano, I-20133 Milano, Italy}
\author{Gianluca Bertaina}
\email{g.bertaina@inrim.it}
\affiliation{Istituto Nazionale di Ricerca Metrologica, Strada delle Cacce 91, I-10135 Torino, Italy}
\author{Davide E. Galli}
\email{davide.galli@unimi.it}
\affiliation{Dipartimento di Fisica ``Aldo Pontremoli'', Universit\`a degli Studi di Milano, via Celoria 16, I-20133 Milano, Italy}

\begin{abstract}
We introduce a method to simulate open quantum many-body dynamics by combining time-dependent variational Monte Carlo (tVMC) with quantum trajectory techniques. Our approach unravels the Lindblad master equation into an ensemble of stochastic Schrödinger equations for a variational ansatz, avoiding the exponential cost of density matrix evolution. 
The method is compatible with generic ans\"{a}tze, including expressive neural-network wavefunctions.
We derive the nonlinear stochastic equations of motion for the variational parameters and employ suitable Stratonovich numerical solvers. To validate our approach, we simulate quenches in the locally dissipative long-range Ising model in a transverse field, accurately capturing non-equilibrium magnetization and spin squeezing dynamics relevant to trapped-ion and Rydberg atom experiments. The framework is computationally efficient, scalable on high-performance computing platforms, and can be readily integrated into existing tVMC implementations. This work enables large-scale simulations of complex, dissipative quantum systems in higher dimensions, with broad implications for quantum technology and fundamental science.
\end{abstract}

\maketitle
Understanding and controlling the dynamics of open quantum many-body systems
~\cite{Breuer_TheoryOpenQuantum_2002,Vacchini_OQS_2024} is central to the development of quantum technologies. The roles of dissipation and decoherence have been extensively investigated across a range of areas, including quantum computation~\cite{Cai_Quantumerrormitigation_2023}, quantum metrology~\cite{Pezze_Quantummetrologynonclassical_2018,MONTENEGRO20251}, quantum communication~\cite{Azuma_Quantumrepeatersquantum_2023} and quantum thermodynamics~\cite{Campaioli_ColloquiumQuantumbatteries_2024}.
In fact, the interplay of environment-induced noise with many-body coherence and correlations in driven, out-of-equilibrium regimes offers a fertile ground for investigating fundamental aspects of quantum physics, with direct implications for the study of thermalization, quantum coherence, entanglement, dissipative phase transitions and information flow in noisy environments~\cite{Sieberer_DynamicalCriticalPhenomena_2013,Fitzpatrick_ObservationDissipativePhase_2017,Song_Crossoverdiscontinuouscontinuous_2023,Ranabhat_Thermalizationlongrange_2024,agarwal1977,carmichael1977,morrison2008,Hannukainen2018,iemini2018,Ferioli2023,Goncalves2025}. 
These investigations also pave the way for technological applications, particularly in the emerging domain of critical quantum metrology~\cite{PorrasPRA_2017,Heugel_2019,DiCandia_2023,Garbe_2020,Ilias_2022,Montenegro2023,Pavlov_2023,Cabot2024}.
A wide range of platforms, such as trapped ions, Rydberg atoms, ultracold gases, and superconducting circuits, has made the controlled realization of open quantum dynamics experimentally accessible, prompting the development of powerful theoretical tools. 

Due to the high dimensionality of the density matrix for quantum many-body systems, various approaches and numerical techniques have been devised to simulate Markovian master equations~\cite{Weimer_2021}, which are particularly efficient in the semiclassical regime~\cite{Plankensteiner_QuantumCumulantsjlJulia_2022,Huber_Realisticsimulationsspin_2022} or employ variational techniques~\cite{Bond_Openquantumdynamics_2024}, and neural-network representations of the density matrix~\cite{Hartmann_2019,Nagy_VariationalQuantumMonte_2019,Vicentini_VariationalNeuralNetworkAnsatz_2019}, trained via stochastic minimization of the Lindbladian residual. Matrix product state techniques are also an intense field of study and are accurate in the absence of entanglement volume law~\cite{Mascarenhas_Matrixproductoperatorapproachnonequilibrium_2015,Sulz_Numericalsimulationlongrange_2024,Hryniuk_TensornetworkbasedvariationalMonte_2024,qbfkzrzc}. Complementarily, open libraries are available for exact benchmarks with relatively small systems~\cite{JOHANSSON20121760,JOHANSSON20131234,Shammah_Openquantumsystems_2018,Mercurio_QuantumToolboxjlefficientJulia_2025}.

In this work, we introduce a scalable and flexible variational scheme for simulating open quantum many-body dynamics described by a Markovian master equation in the Lindblad form, by combining the time-dependent variational Monte Carlo (tVMC) framework~\cite{Carleo_LocalizationGlassyDynamics_2012,Carleo_Becca_Sanchez_2014,carleo_solving_2017,Schmitt_Quantummanybodydynamics_2020,Salioni_Adaptivequantumdynamics_2025} with the quantum trajectory formalism~\cite{Gisin_quantumstatediffusionmodel_1992,Dalibard1992}. This hybrid approach reformulates the Lindblad evolution as an ensemble of stochastic pure-state trajectories, each evolved with a variationally optimal dynamics. It allows us to retain the expressive power and scalability of tVMC while incorporating dissipation nonperturbatively and without uncontrolled approximations. Access to the full time evolution and not only to the stationary state is crucial for quantum technologies, in particular for quantum state engineering and quantum metrology protocols.
In this respect, we apply our method to the long-range Ising model in a transverse field under local dissipation~\cite{Koffel_EntanglementEntropyLongRange_2012,Pagano_Quantumapproximateoptimization_2020,Paz_DrivendissipativeIsingmodel_2021}, a paradigmatic setting for investigating entanglement generation. We analyze the emergence of metrologically useful spin squeezing during quench dynamics~\cite{Kitagawa_Squeezedspinstates_1993,Wineland_Spinsqueezingreduced_1992,Wineland_Squeezedatomicstates_1994,Ma_Quantumspinsqueezing_2011,Pezze_Quantummetrologynonclassical_2018,Comparin_Robustspinsqueezing_2022} in regimes where strong correlations, long-range interactions~\cite{Defenu_Outofequilibriumdynamicsquantum_2024}, and noise coexist, highlighting the promise of variational trajectory-based approaches for simulating complex open-system dynamics beyond the reach of existing techniques.

\textit{Problem and approach} --- When the system interacts with an environment that retains no memory of the past evolution, the system's density matrix $\hat \rho$ undergoes Markovian dynamics, governed by the Lindblad master equation~\cite{Breuer_TheoryOpenQuantum_2002},
\begin{equation}\label{eq:Lindblad}
 \frac{d \hat \rho}{dt}
 = -\frac{i}{\hbar} \left[ \hat H, \hat \rho \right] + \sum_n \left( \hat{L}_n^{} \hat \rho \hat{L}_n^\dagger - \frac{1}{2} \left\lbrace \hat{L}_n^\dagger \hat{L}_n^{}, \hat \rho \right\rbrace \right) \,\text{,}
\end{equation}
where $\hat H$ is the system Hamiltonian and $\hat{L}_n$ are the collapse operators that describe the dissipative channels due to the system-environment interaction. Our aim is to compute the expectation value of a general system observable $\hat A$ at time $t$, which we denote by $\llangle \hat A \rrangle = \operatorname{Tr}\left( \hat A \hat \rho \right)$.

If the Hilbert space of the system has dimension $d$, then computing $\llangle \hat A \rrangle$ from the full density matrix entails a computational cost of at least $O(d^2)$, due to the $O(d^2)$ independent entries in $\hat \rho$. Alternatively, the density matrix can be represented as an ensemble average over pure states, allowing $\llangle \hat A \rrangle$ to be estimated from the average over pure-state expectation values. To this end, we employ the quantum state diffusion approach~\cite{Gisin_quantumstatediffusionmodel_1992}, in which an ensemble of pure-state \textit{quantum trajectories} evolves according to an It\^o stochastic differential equation (SDE):
\begin{multline}\label{eq:SSE_nonlin_Ito}
 d | \psi \rangle =\\
 \left\lbrace \left[ - \frac{i}{\hbar} \hat H - \frac{1}{2} \sum_n \left( \hat{L}_n^\dagger \hat{L}_n^{} - \langle \hat{L}_n^{} + \hat{L}_n^\dagger \rangle \hat{L}_n^{} + \frac{1}{4} \langle \hat{L}_n^{} + \hat{L}_n^\dagger \rangle^2 \right) \right] dt \right. \\
 \left. + \sum_n \left( \hat{L}_n^{} - \frac{1}{2} \langle \hat{L}_n^{} + \hat{L}_n^\dagger \rangle \right) dW_n^{} \right\rbrace | \psi \rangle \,\text{,}
\end{multline}
where $| \psi \rangle$ is the pure state of a single trajectory at time $t$, the $dW_n$ are independent Wiener increments with variance $dt$, and $\langle \bullet \rangle$ denotes an expectation value over the state $|\psi\rangle$. Eq.~\eqref{eq:SSE_nonlin_Ito}, known as the stochastic Schr\"{o}dinger equation (SSE), is nonlinear in $|\psi\rangle$, since there are terms depending on expectation values. This particular form of the SSE preserves the norm of $|\psi\rangle$ throughout the evolution.

If the initial density matrix is given by $\hat \rho (0) = \sum_i p_i | \psi_{0,i} \rangle \langle \psi_{0,i} |$, where $p_i$ are classical probabilities and $| \psi_{0,i} \rangle$ are normalized pure states, then each quantum trajectory is initialized in the state $| \psi_{0,i} \rangle$ with probability $p_i$. The subsequent evolution of each trajectory follows the SSE~\eqref{eq:SSE_nonlin_Ito}. After generating $N_\text{T}$ such trajectories, denoted by $\lbrace |\psi_j\rangle \rbrace_{j=1,\dots,N_\text{T}}$, the expectation value of the observable $\hat A$ is obtained by ensemble averaging,
\begin{equation}\label{eq:traj_average}
 \llangle \hat A \rrangle = \frac{1}{N_\text{T}} \sum_j \langle \psi_j | \hat A | \psi_j \rangle \,\text{.}
\end{equation}

This trajectory-based calculation has a computational complexity of $O(N_\text{T} d)$. 
For example, choosing $N_\text{T} \approx 1000$ (which typically reduces the statistical error to a few percent) yields a complexity significantly lower than $O(d^2)$, even for systems of moderate size.
Furthermore, since each trajectory is independent, the calculation is trivially parallelizable, further reducing the overall computation time.

Despite these advantages, the Hilbert space dimension $d$ still grows exponentially with the system size. To address this difficulty, we approximate the quantum state of each trajectory using a variational ansatz of the form $e^{\phi} | \psi_{\boldsymbol{\theta}}^{} \rangle$, where $\phi \in \mathbb{C}$ is a complex parameter controlling the global phase and norm, and $\boldsymbol{\theta} = (\theta_1, \dots, \theta_M) \in \mathbb{C}^M$ is a set of $M$ complex variational parameters. Given a configuration $x$ of the system, the corresponding variational wavefunction is expressed as $e^{\phi} \psi_{\boldsymbol{\theta}}(x)^{} = e^{\phi} \langle x| \psi_{\boldsymbol{\theta}}^{} \rangle$. This approach reduces the effective dimensionality of the state space from $d$ to $M+1$, lowering the computational complexity to $O(N_\text{T} M)$. The central task then becomes determining how the variational parameters evolve in time along each quantum trajectory.

\textit{Variational quantum trajectories} --- To derive the equations of motion for the variational parameters, we need to apply the chain rule for differentials. However, in It\^o calculus, the standard chain rule does not hold. Therefore, we reformulate the SSE~\eqref{eq:SSE_nonlin_Ito} in the Stratonovich formalism, where the chain rule holds in its conventional form. Applying the standard It\^o to Stratonovich conversion~\cite{Gardiner_HandbookStochasticMethods_1996}, we obtain
\begin{equation}\label{eq:SSE_nonlin_Strat}
 d|\psi\rangle = \left[ -\frac{i}{\hbar} \hat H^\text{Strat}_\text{eff} \, dt + \sum_n \left( \hat{L}_n^{} - \frac{1}{2} \langle \hat{L}_n^{} + \hat{L}_n^\dagger \rangle\right) \circ dW_n^{} \right] | \psi \rangle \,\text{,}
\end{equation}
where the effective Hamiltonian $\hat H^\text{Strat}_\text{eff}$ in the Stratonovich formulation is given by
\begin{multline}
 \hat H^\text{Strat}_\text{eff}= \hat H - \frac{i \hbar}{2} \sum_n \bigg( \hat{L}_n^\dagger \hat{L}_n^{} - 2 \langle \hat{L}_n^{} + \hat{L}_n^\dagger \rangle \hat{L}_n^{} + \hat{L}_n^2 \\
 + \langle \hat{L}_n^{} + \hat{L}_n^\dagger \rangle^2 - \frac{1}{2} \langle \hat{L}_n^2 + 2 \hat{L}^\dagger_n \hat{L}_n^{} + \hat{L}_n^{\dagger 2} \rangle \bigg) \,\text{.}
\end{multline}
As usual, the symbol $\circ \, dW_n$ denotes a Stratonovich differential.

We are now in a position to address the problem of approximating the evolution described by Eq.~\eqref{eq:SSE_nonlin_Strat} using a generic variational ansatz of the form $e^{\phi} | \psi_{\boldsymbol{\theta}}^{} \rangle$. Working within the Stratonovich formalism allows us to apply the standard chain rule for differentials, yielding
\begin{equation}\label{eq:chain_differential}
 d\left(e^{\phi} | \psi_{\boldsymbol{\theta}}^{} \rangle \right) = \sum_k e^\phi \frac{\partial}{\partial \theta_k} | \psi_{\boldsymbol{\theta}}^{} \rangle d\theta_k + e^\phi | \psi_{\boldsymbol{\theta}} \rangle d\phi \, \text{.}
\end{equation}
To derive the equations of motion for the variational parameters, we minimize the Hilbert space distance between the exact SSE increment~\eqref{eq:SSE_nonlin_Strat} and the variational increment~\eqref{eq:chain_differential} at each time~\cite{Carleo_LocalizationGlassyDynamics_2012,Carleo_Becca_Sanchez_2014}:
\begin{equation}
 \begin{aligned}
  \Delta^2 =& \left\| d | \psi \rangle - d \left(e^{\phi} | \psi_{\boldsymbol{\theta}}^{} \rangle \right) \right\|^2 \\
  =& \left\| d | \psi \rangle - \sum_k e^\phi \frac{\partial}{\partial \theta_k} | \psi_{\boldsymbol{\theta}}^{} \rangle d\theta_k + e^\phi | \psi_{\boldsymbol{\theta}} \rangle d\phi \right\|^2 \,\text{.}
 \end{aligned}
\end{equation}
Minimizing $\Delta^2$ with respect to the stochastic differentials $d\theta_k$ and $d\phi$ yields a set of Stratonovich SDEs for the variational parameters $\theta_k$:
\begin{equation}\label{eq:otVMC_nonlin}
 \sum_{k'} S_{k,k'}^{} \, d \theta_{k'}^{} = - \frac{i}{\hbar} F_k^{} dt + \sum_n N_k^{n} \circ dW_n^{} \;,
\end{equation}
for $k=1,\dots,M$, where, by introducing the \textit{local operators} $\mathcal{O}_k(x) = \frac{\partial}{\partial \theta_k} \ln \langle x | \psi_{\boldsymbol{\theta}}^{} \rangle$,
the \textit{local effective energy} $\mathcal{E}_\text{eff}(x) = \langle x | \hat H^\text{Strat}_\text{eff} | \psi_{\boldsymbol{\theta}}^{} \rangle / \langle x | \psi_{\boldsymbol{\theta}}^{} \rangle$,
and the \textit{local noises} $\mathcal{B}_{}^{n}(x) = \langle x | \hat{L}_n^{} | \psi_{\boldsymbol{\theta}}^{} \rangle / \langle x | \psi_{\boldsymbol{\theta}}^{} \rangle$, the quantum geometric tensor $S$ and the quantum force $F$ and noise $N^n$ vectors are defined as follows: 
$S_{k,k'}^{} = \langle \mathcal{O}_k^* \mathcal{O}_{k'}^{} \rangle - \langle \mathcal{O}_k^* \rangle \langle \mathcal{O}_{k'}^{} \rangle$,
$F_{k}^{} = \langle \mathcal{O}_k^* \mathcal{E}_\text{eff} \rangle - \langle \mathcal{O}_k^* \rangle \langle \mathcal{E}_\text{eff} \rangle$, and
$N_{k}^{n} = \langle \mathcal{O}_k^* \mathcal{B}^{n} \rangle - \langle \mathcal{O}_k^* \rangle \langle \mathcal{B}^{n} \rangle$. With a common abuse of notation, here we indicate the Monte Carlo average of a local variable $\mathcal{A}$ with $\langle\mathcal{A}\rangle$.

Minimizing $\Delta^2$ also yields a SDE for the parameter $\phi$,
\begin{multline}\label{eq:dphi_nonlin}
d\phi = - \sum_k \langle \mathcal{O}_k \rangle d\theta_k - \frac{i}{\hbar} \langle \mathcal{E}_\text{eff} \rangle dt \\
+ \sum_n \left( \langle \mathcal{B}^{n} \rangle - \frac{1}{2} \langle \hat L_n + \hat L_n^\dagger \rangle \right) \circ dW_n \,\text{,}
\end{multline}
that depends on the solutions $d\theta_k$ of Eq.~\eqref{eq:otVMC_nonlin}. Although this equation can be integrated to track the evolution of $\phi$, it is not necessary to do so in practice, since $\phi$ encodes only gauge degrees of freedom, namely, the global phase and norm, which do not affect the physical information encoded in the state vector. Nevertheless, it plays an essential auxiliary role in the derivation of the variational equations of motion.

The system of equations~\eqref{eq:otVMC_nonlin} governs the evolution of the variational parameters $\boldsymbol{\theta}$ and lies at the core of our method, which we refer to as the open time-dependent variational Monte Carlo (otVMC). An otVMC simulation proceeds as follows. As is commonplace in the quantum diffusion approach, we assume a pure initial state. Within our variational representation, it is identified by a parameter vector $\boldsymbol{\theta}_0$, such that the initial density matrix is $\hat \rho(0) \propto |\psi_{\boldsymbol{\theta}_0}\rangle \langle \psi_{\boldsymbol{\theta}_0} |$. We then generate a set of $N_\text{T}$ quantum trajectories, each initialized from the state $|\psi_{\boldsymbol{\theta}_0} \rangle$.

For a given trajectory at time $t$, with variational parameters $\boldsymbol{\theta}(t)$, system configurations $x$ are sampled using the Metropolis algorithm from the normalized probability distribution obtained by the square modulus $|\psi_{\boldsymbol{\theta}(t)} (x)|^2 = | \langle x | \psi_{\boldsymbol{\theta}(t)}\rangle |^2$. These samples are used to compute Monte Carlo estimates of the expectation values required to evaluate the $S$ matrix and the $F$ and $N^n$ vectors appearing in Eq.~\eqref{eq:otVMC_nonlin}. The resulting SDE is then integrated using a numerical scheme with time step $\Delta t$, yielding the updated parameters $\boldsymbol{\theta}(t+\Delta t)$. 
To this aim, we employ a regularization based on the signal-to-noise ratio (SNR) of the components of the vector $D = - (i/\hbar) F^{} \Delta t + \sum_n N^{n} \Delta W_n^{}$, where $\Delta W_n$ are the discretized Wiener increments; the vector $D$ is expressed in the basis that diagonalizes $S$, and its noisy components are suppressed following the regularization scheme described in the Supplemental material of Ref.~\cite{Schmitt_Quantummanybodydynamics_2020}, with an SNR threshold of $4$.
Particular care must be taken in employing a scheme consistent with the Stratonovich interpretation. We employ a midpoint predictor-corrector solver (see App.~\ref{app:solver} for details). This process is repeated iteratively for the desired number of time steps to simulate the full trajectory.

\textit{Application to the LITF model} ---
To demonstrate the capabilities of the otVMC method, we tackle a challenging model, the dissipative long-range Ising model in a transverse field (LITF)~\cite{Koffel_EntanglementEntropyLongRange_2012,Morrison_DynamicalQuantumPhase_2008}. Experimental realizations, particularly with trapped ions \cite{Monroe_Quantumsimulationspin_2015, Bohnet_Quantumspindynamics_2016}, have demonstrated the practical relevance of this model. Recent experiments have explored critical behavior through the Kibble-Zurek mechanism \cite{Li_ProbingCriticalBehavior_2023} and investigated optimization protocols \cite{Pagano_Quantumapproximateoptimization_2020}. The transverse-field one-axis-twisting (OAT) regime has been particularly well-studied experimentally \cite{Borish_TransverseFieldIsingDynamics_2020}. In one dimension (1D) and for $N$ two-level quantum systems, the Hamiltonian of the LITF model is 
\begin{equation}
\hat{H} = -\frac{J}{\mathcal{K}(\alpha)}\sum_{i<j} \frac{\hat{\sigma}^z_i \hat{\sigma}^z_j}{|i-j|^\alpha} - h \sum_i \hat{\sigma}^x_i
\end{equation}
where the ferromagnetic coupling $J>0$ and the transverse field $h$ compete and favor a ferromagnetic or paramagnetic state along the $x$ direction, respectively. The dissipative part of the model is exemplified by local decay Lindblad operators $\hat{L}_i=\sqrt{\kappa}\hat{\sigma}_i^-$ for $i=1\cdots N$. 
The power-law exponent $\alpha$ allows for interpolation between the two extremes: the standard nearest-neighbor Ising model ($\alpha=\infty$), and the OAT Hamiltonian, or isotropic Lipkin-Meshkov-Glick model ($\alpha=0$). We focus on the intriguing marginal case $\alpha=1$ in 1D.
The Kac factor $\mathcal{K}$ is introduced to guarantee extensive energies, and is defined as $\mathcal{K}(\alpha)=\sum_{i<j} |i-j|^{-\alpha}/(N-1)$. 
We consider periodic boundary conditions and include interactions only with the nearest replicas.
When $h=0$, this model can be treated by the techniques described in Refs.~\cite{FossFeig_Nonequilibriumdynamicsarbitraryrange_2013,McDonald_ExactSolutionsInteracting_2022}, while the general case with $\alpha=0$ has been recently solved in the steady state~\cite{Roberts_ExactSolutionInfiniteRange_2023}. An interesting and useful property of the LITF model is its ability to generate entanglement in the form of spin squeezing during a quench from the paramagnetic state~\cite{Comparin_Robustspinsqueezing_2022}. We then model the quench dynamics starting from a coherent spin state along the $+x$ direction. The coordinates are labeled as $\bm{\sigma}=\{\sigma_1,\sigma_2,\cdots,\sigma_N\}$, where $\sigma_i=\pm 1$ are the eigenvalues of the local $\hat{\sigma}^z_i$ operator. This choice of computational basis prevents possible sampling problems~\cite{Sinibaldi_2023}. 

A few observables are of particular interest. The magnetization $M_i$, for $i=x,y,z$, is the expectation value of the normalized collective spin vector, $\hat{M}_i=(1/N)\sum_a^N \hat{\sigma}_a^i$. The normalized spin covariance matrix $C_{ij}=(1/2)\langle\{\hat{M}_i,\hat{M}_j\}\rangle-M_i M_j$ describes the fluctuations around the mean collective spin.
Wineland's metrological spin-squeezing parameter~\cite{Wineland_Spinsqueezingreduced_1992,Wineland_Squeezedatomicstates_1994} quantifies the minimal fluctuation of the spin perpendicular to the average spin, normalized to the magnetization, $\xi^2=N {\min_{\perp} \Delta^2 M_\perp}/{|\bar{M}|^2}$, which we evaluate as described in Ref.~\cite{Caprotti_Analysisspinsqueezinggeneration_2024}. 

\begin{figure}[tbph]
 \centering
 \includegraphics[width=\columnwidth]{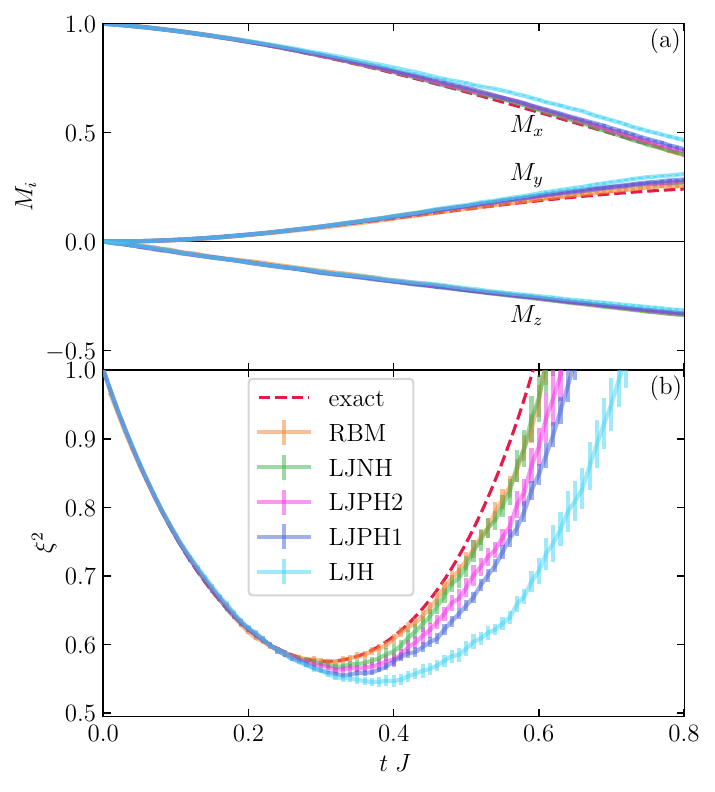}
 \caption{Evolution of magnetization components (panel a) and spin squeezing parameter (panel b) for $N=10$, $\kappa=0.5J$ and $h=0$. Comparison of different ans\"{a}tze (solid lines) to the exact result of App.~\ref{app:fossfeig} (dashed line). In panel a of all figures, uncertainty bars are smaller than symbol sizes.}
 \label{fig:N10}
\end{figure}

We investigate the accuracy and computational cost of various variational ans\"{a}tze: a very expressive neural network quantum state of restricted Boltzmann machine (RBM) type~\cite{carleo_solving_2017}; and a less expensive long-range Jastrow (LJ) wavefunction with single and two-body correlations with different levels of non-uniformity.
The RBM state is defined by:
\begin{equation}\label{eq:RBM}
 \psi^{\text{RBM}}_{\bm{\theta}}(\bm{\sigma})=\exp{\left[ \sum_{i=1}^N a_i\sigma_i \right]} \times \prod_{j=1}^{N^\prime} 2\cosh{\gamma_j(\bm{\sigma})}\;,
\end{equation}
with $\gamma_j(\bm{\sigma}) = b_j + \sum_{i=1}^N w_{ij}\sigma_i$. The total number of variational parameters is $N N^\prime+N+N^\prime$ and we consider a number of hidden units $N^\prime=N$. The initial paramagnetic state is exactly parametrized by $b=a=w=0$; since the quantum force $F$ is null at that point, special care has to be taken in bootstrapping tVMC~\cite{Apostoli2025order2}. The LJ state is defined by:
\begin{equation}\label{eq:LJ}
 \psi^{\text{LJ}}_{\bm{\theta}}(\bm{\sigma}) = \exp\left[{\sum_{i=1}^N a_i\sigma_i}\right] \times \exp{ \sum_{j<k}\eta_{jk}\,\sigma_j\sigma_k }~.
\end{equation}
While in the totally inhomogeneous case (LJNH) this ansatz requires $N(N+1)/2$ parameters (in the open boundary conditions case), comparable to the RBM, constraints can be imposed that improve scalability at the cost of slightly reducing accuracy. The partially inhomogeneous variant (LJPHd) sets all two-body correlations $\eta_{jk}$ to be dependent only on the distance $D=|j-k|$ for all $D>d$ where $d$ can be reduced from $N-1$ (corresponding to LJNH) to 0, corresponding to the homogeneous case (LJH). To accommodate the local action of noise, we always consider the biases $a_i$ to be inhomogeneous. Therefore LJPHd has $2N-1-d +N d/2$ parameters. The initial paramagnetic state is obtained by setting all parameters equal to zero. 

\begin{figure}[tbph]
 \centering
 \includegraphics[width=0.99\columnwidth]{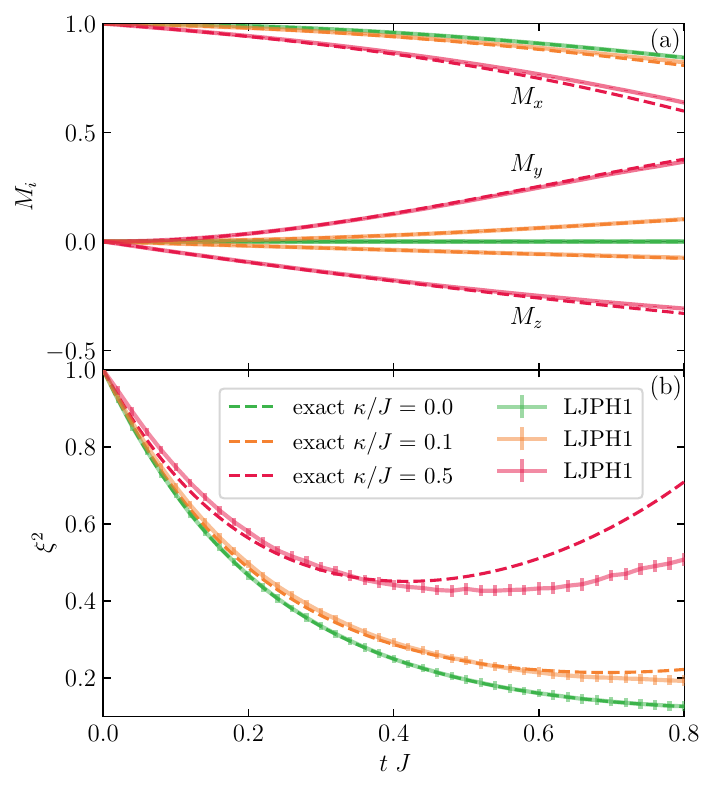}
 \caption{Study of magnetization and spin squeezing for $N=160$ employing the LJPH1 ansatz, with varying $\kappa/J=0, 0.1, 0.5$ and no transverse field (solid lines), compared to the exact result of App.~\ref{app:fossfeig} (dashed lines).}
 \label{fig:kappa}
\end{figure}

In Fig.~\ref{fig:N10}, we show the time evolution of the components of magnetization (panel a) and the spin-squeezing parameter (panel b) for $N=10$ spins, $\kappa=0.5 J$ and no transverse field. We benchmark the otVMC results with different ans\"{a}tze to the exact result from Ref.~\cite{FossFeig_Nonequilibriumdynamicsarbitraryrange_2013} (see App.~\ref{app:fossfeig}). All results are obtained by averaging over $N_\text{T}=400$ trajectories. We notice that all employed wavefunctions yield accurate results for short times, in particular for the magnetization along $z$. This is probably a merit of the inhomogeneous bias parameters. The RBM and LJNH wavefunctions are highly consistent with each other and are very accurate even for the spin-squeezing parameter around its minimum. The computationally cheaper LJPHd ansatz performs very well for magnetization and well for the challenging $\xi^2$ around its minimum, provided that $d>0$. In particular, while LJH does not sufficiently accommodate inhomogeneity, we argue that LJPH1 strikes a good compromise between scalability and accuracy. See App.~\ref{app:cost} for considerations on the computational cost of our simulations. 

In Fig.~\ref{fig:kappa}, we demonstrate that our method with the LJPH1 ansatz easily scales to very large particle numbers, relevant to current experiments. We focus on a large $N=160$ spin system and study the evolution of observables with varying $\kappa$ in the absence of transverse field. For small $\kappa$, the LJPH1 results are remarkably accurate, with a slight overestimation of the minimum of $\xi^2$, with increasing $\kappa$. 
Larger values of $\kappa$ (not shown) induce larger jumps in the variational parameters and sometimes require higher tVMC sampling than usual.

\begin{figure}[tbph]
 \centering
 \includegraphics[width=0.99\columnwidth]{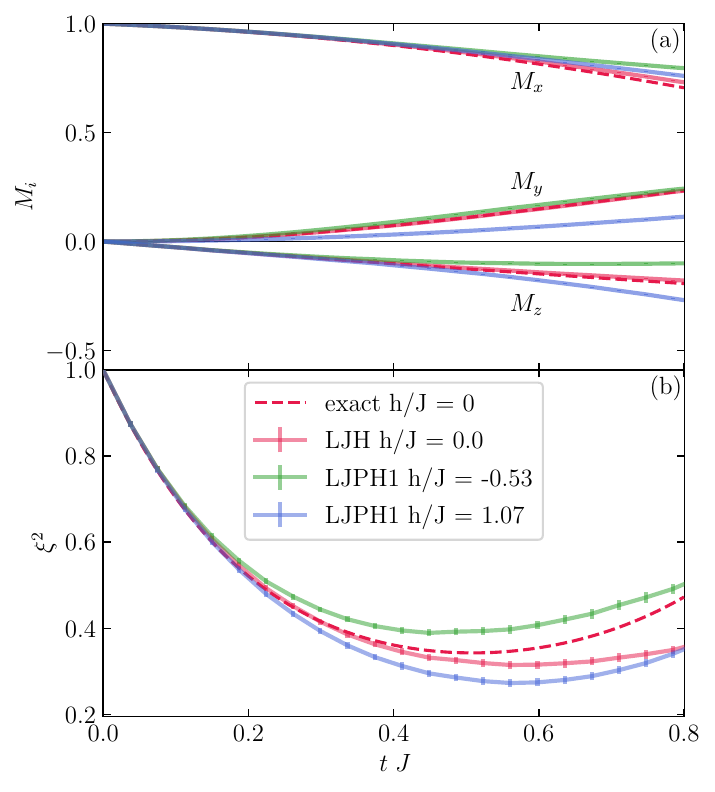}
 \caption{Study of magnetization and spin squeezing for $N=100$ and $\kappa/J=0.27$ with varying transverse field $h=0, -0.53, 1.07$. We employ the LJPH1 and LJH ans\"{a}tze (solid lines) and report the exact result of App.~\ref{app:fossfeig} for $h=0$ (dashed lines).}
 \label{fig:transverse}
\end{figure}

Finally, in Fig.~\ref{fig:transverse}, we focus on $N=100$ and investigate the impact of transverse magnetic field in the dissipative LITF model, a system for which there is no analytical solution, to our knowledge. It has been argued that inclusion of a transverse field provides spin-squeezing even in the stationary state, in the OAT case~\cite{Lee_DissipativetransversefieldIsing_2013}, and, in our simulations with $\alpha=1$, we observe that there is a noticeable impact also close to the minimum of $\xi^2$ during a quench, depending on the sign of $h$. $h>0$ proves to be advantageous, while $h<0$ reduces spin-squeezing generation. One can find a heuristic explanation of this effect by considering the tilted OAT Hamiltonian $-(J N/2) \hat{M}_z^2 -h N \hat{M}_x$. Looking at the $yz$ plane tangent to the initial state in a Bloch sphere representation, the quadratic term generates a squeezed spin distribution along the $z=-y$ axis, while the dissipation term translates it to the lower hemisphere, thereby enhancing the effective rotation towards positive $y$. A transverse field with $h<0$ rotates the squeezed distribution towards the equator, preventing further squeezing, while $h>0$ brings the distribution closer to alignment with the $z$ axis, dynamically allowing the quadratic term to have a greater effect.

\textit{Conclusions} ---
We have shown an efficient and scalable time-dependent variational Monte Carlo method to simulate the Lindblad dynamics of many-body open quantum systems. Our approach leverages the unraveling of the master equation into stochastic Schr\"{o}dinger trajectories within a compact yet expressive variational manifold, exemplified here by different ans\"{a}tze.
We have focused on a one-dimensional long-range model which is highly non-trivial and relevant to experiments, considering the case without a transverse magnetic field, for benchmarking purposes, and the transverse field case, to demonstrate the flexibility of the method. Remarkably, our approach can scale to hundreds of particles, owing to its natural parallelizability, and is well suited to integration into existing codebases. It can access not only the stationary state, but the full open quantum dynamics, which is important for modeling quantum technologies.
Our method can be applied to more complex models in higher dimensions, where tensor network approaches are not guaranteed to be accurate in general. 
Other relevant research directions are the extension to dissipative systems in the continuum, like ultracold gases and liquids, to quantum optics systems such as cavity-coupled atoms, and to the study of defect formation in quenches~\cite{Li_ProbingCriticalBehavior_2023}. Suitable sampling will also give access to mixed initial states, such as thermal ones~\cite{GonzalezLazo_Finitetemperaturecriticalbehavior_2021,Ranabhat_Thermalizationlongrange_2024}. 
Our solution to the dynamics of nonlinear trajectories will also provide a route to model dissipative systems under continuous measurement~\cite{Albarelli_pedagogicalintroductioncontinuously_2024,Caprotti_Analysisspinsqueezinggeneration_2024}.

The data that support the findings of this article are openly available~\cite{apostoli_zenodo}.

\acknowledgments
We thank T. Comparin for discussing with us the idea that fostered this work. We acknowledge stimulating discussions with T. Roscilde and G. Carleo. We acknowledge the early contributions of L. D'Onofrio and S. Lonardoni, developed in the framework of their theses.
We acknowledge funding from Italian Ministry of Research and Next Generation EU via the PRIN 2022 Project CONTRABASS (Contract No. 2022KB2JJM). 
We acknowledge the CINECA awards IsCb5-PSIOQUAS and IsCc7-SDOQS under the ISCRA initiative, for the availability of high-performance computing resources and support. 
We acknowledge computational resources provided by INDACO Platform, which is a project of high performance computing
at Università degli Studi di Milano.


%

\onecolumngrid
\newpage
\begin{center}
 \textbf{\large End Matter}
\end{center}
\twocolumngrid

\appendix

\section{Alternative derivation of the otVMC equations}
The derivation of the Stratonovich SSE~\eqref{eq:SSE_nonlin_Strat} from the It\^o SSE~\eqref{eq:SSE_nonlin_Ito} is complicated by the presence of a nonlinear term, $-\frac{1}{2}\langle \hat{L}_n^{} + \hat{L}_n^\dagger \rangle$, which multiplies the Wiener increment $dW_n^{}$. To simplify this step and facilitate the derivation of the otVMC equations, one can perform a change of variables, introducing a non-normalized state vector
\begin{equation}\label{eq:change_variables}
 | \tilde \psi(t) \rangle = e^{X(t)} | \psi(t) \rangle \, \text{,}
\end{equation}
where $X(t)$ is a real-valued stochastic process satisfying the following It\^o SDE:
\begin{multline}
 dX = \mu \, dt + \sum_n \sigma_n^{} \, dW_n^{} \,\text{,} \\
 \text{with } \sigma_n^{} = \frac{1}{2} \langle \hat{L}_n^{} + \hat{L}_n^\dagger \rangle \text{, } \mu = \sum_n \sigma_n^2 \,\text{.}
\end{multline}
Using these definitions to derive the It\^o SSE for $| \tilde \psi \rangle$, one finds
\begin{multline}\label{eq:SSE_nonlin_ItonoW}
 d | \tilde\psi \rangle = 
 \left\lbrace \left[ - \frac{i}{\hbar} \hat H - \frac{1}{2} \sum_n \left( \hat{L}_n^\dagger \hat{L}_n^{} - 2 \langle \hat{L}_n^{} + \hat{L}_n^\dagger \rangle \hat{L}_n^{} \right) \right] dt \right. \\
 \left. + \sum_n \hat{L}_n^{} \, dW_n^{} \right\rbrace | \tilde \psi \rangle \,\text{,}
\end{multline}
where the nonlinear term proportional to $dW_n^{}$ in the original SSE is eliminated, making the diffusion term linear in $| \tilde \psi \rangle$. This transformation greatly simplifies the transition to the Stratonovich formulation, ultimately yielding the following Stratonovich SSE for the non-normalized state vector:
\begin{equation}\label{eq:SSE_nonlin_Strat_trick}
 \begin{aligned}
 d | \tilde \psi \rangle = \left\lbrace -\frac{i}{\hbar} \hat{\tilde{H}}^\text{Strat}_\text{eff} \, dt + \sum_n \hat{L}_n^{} \circ dW_n^{} \right\rbrace | \tilde \psi \rangle \,\text{,}
 \end{aligned}
\end{equation}
where we introduced an effective Hamiltonian $\hat{\tilde{H}}^\text{Strat}_\text{eff} = \hat H - \frac{i \hbar}{2} \sum_n \left( \hat{L}_n^\dagger \hat{L}_n^{} - 2 \langle \hat{L}_n^{} + \hat{L}_n^\dagger \rangle \hat{L}_n^{} + \hat{L}_n^2 \right)$.
The term proportional to $\hat{L}_n^2$ arises from the It\^o-to-Stratonovich conversion~\cite{Gardiner_HandbookStochasticMethods_1996}. It is important to note that the expectation values $\langle \hat{L}_n^{} + \hat{L}_n^\dagger \rangle$ in Eq.~\eqref{eq:SSE_nonlin_Strat_trick} are to be evaluated with respect to the normalized state, i.e., $\langle \hat{L}_n^{} + \hat{L}_n^\dagger \rangle = \langle \psi | \hat{L}_n^{} + \hat{L}_n^\dagger | \psi \rangle = \langle \tilde \psi | \hat{L}_n^{} + \hat{L}_n^\dagger | \tilde \psi \rangle / \langle \tilde \psi | \tilde \psi \rangle$.

We now introduce the variational ansatz $e^\phi |\psi_{\boldsymbol{\theta}} \rangle$, and minimize the Hilbert-space distance between the exact SSE increment~\eqref{eq:SSE_nonlin_Strat_trick} and the variational increment~\eqref{eq:chain_differential}. This procedure yields the same set of evolution equations for the variational parameters $\boldsymbol{\theta}$ as derived in the main text, namely Eq.~\eqref{eq:otVMC_nonlin}. This outcome is expected, as the change of variables in Eq.~\eqref{eq:change_variables} only affects the norm of the state vector, without altering its ray. As a result, the dynamics of $\boldsymbol{\theta}$ remain unchanged.
The parameter $\phi$, on the other hand, evolves according to $d\phi = - \sum_k \langle \mathcal{O}_k \rangle \, d\theta_k - \frac{i}{\hbar} \langle \mathcal{E}_\text{eff} \rangle \, dt + \sum_n \langle \mathcal{B}^n \rangle \circ dW_n$. Since $\phi$ encodes only gauge degrees of freedom, it does not influence the physical predictions of the method.

\section{Linear stochastic Schr\"{o}dinger equation}\label{app:linear}
There also exists an alternative quantum state diffusion approach, that employs a \textit{linear} It\^o SSE~\cite{DAlberto_CombiningQuantumTrajectories_2024},
\begin{equation}\label{eq:SSE_lin_Ito}
 d | \psi \rangle =
 \left\lbrace \left( -\frac{i}{\hbar} \hat H - \frac{1}{2} \sum_n \hat{L}_n^\dagger \hat{L}_n^{} \right) dt + \sum_n \hat{L}_n^{} \, dW_n^{} \right\rbrace | \psi \rangle \,\text{.}
\end{equation}
This equation generates a different set of quantum trajectories, that, in contrast to the nonlinear SSE~\eqref{eq:SSE_nonlin_Ito}, do not preserve the norm of the state $| \psi \rangle$. It is important to stress that, if the trajectories $|\psi_j\rangle$ are generated according to the linear SSE~\eqref{eq:SSE_lin_Ito}, the pure-state expectation values appearing on the right-hand side of Eq.~\eqref{eq:traj_average} are not normalized, but the formula still holds as it is~\cite{Albarelli_pedagogicalintroductioncontinuously_2024}.

Like for the nonlinear SSE, it is possible to use a variational ansatz $e^\phi |\psi_{\boldsymbol{\theta}} \rangle$ to approximate the quantum trajectories generated by the linear SSE. The Stratonovich form of the equation is given by
\begin{equation}\label{eq:SSE_lin_Strat}
 d|\psi\rangle =
 \left\lbrace -\frac{i}{\hbar} \hat H^{\text{Strat,lin}}_\text{eff} + \sum_n \hat{L}_n^{} \circ dW_n^{} \right\rbrace | \psi \rangle \,\text{.}
\end{equation}
where we defined an appropriate effective Hamiltonian $\hat H^{\text{Strat,lin}}_\text{eff} = \hat H - \frac{i}{2} \sum_n \left( \hat{L}_n^\dagger \hat{L}_n^{} + \hat{L}_n^2 \right)$.
An analogous distance-minimization procedure as the one carried out for the nonlinear SSE yields a set of equations of motion for $\boldsymbol{\theta}$ that is formally identical to Eq.~\eqref{eq:otVMC_nonlin}, with the substitutions $\mathcal{E}_\text{eff}^{} \rightarrow \mathcal{E}_\text{eff}^\text{lin} = \langle x | \hat H^{\text{Strat,lin}}_\text{eff} | \psi_{\boldsymbol{\theta}}^{} \rangle / \langle x | \psi_{\boldsymbol{\theta}}^{} \rangle$ and, subsequently, $F_k \rightarrow F_k^\text{lin} = \langle \mathcal{O}_k^* \mathcal{E}^\text{lin}_\text{eff} \rangle - \langle \mathcal{O}_k^* \rangle \langle \mathcal{E}^\text{lin}_\text{eff} \rangle$. This implies the relation
\begin{equation}
 F^\text{lin} = F - i \hbar \sum_n \langle \hat{L}_n^{} + \hat{L}_n^\dagger \rangle N^n \text{,}
\end{equation}
which succinctly expresses the difference between the otVMC equations for the linear and nonlinear SSEs. Parameter $\phi$ evolves according to $d\phi = - \sum_k \langle \mathcal{O}_k \rangle d\theta_k - \frac{i}{\hbar} \langle \mathcal{E}^\text{lin}_\text{eff} \rangle dt
+ \sum_n \langle \mathcal{B}^{n} \rangle \circ dW_n$.

In comparison to the nonlinear SSE case, here we have an extra complication: the expectation value of an observable $\hat A$ on the unconditional mixed state is given by the average over trajectories of $\langle \psi | \hat A | \psi \rangle$, where $| \psi \rangle$ is a \textit{non-normalized} state (in contrast to the case of the nonlinear SSE), which is approximated by our variational ansatz $e^{\phi} | \psi_{\boldsymbol{\theta}} \rangle$. However, Monte Carlo integration, employing the Metropolis algorithm, only yields normalized expectation values, i.e., estimations of $\langle \psi_{\boldsymbol{\theta}} | \hat A | \psi_{\boldsymbol{\theta}} \rangle / \langle \psi_{\boldsymbol{\theta}} | \psi_{\boldsymbol{\theta}} \rangle$. To recover the non-normalized expectations, it is necessary to know the value of the normalization $\mathcal{Q}(t) = e^{\phi^*(t) + \phi(t)} \langle \psi_{\boldsymbol{\theta}(t)} | \psi_{\boldsymbol{\theta}(t)} \rangle$. To this end, we compute the stochastic differential of $\mathcal{Q}(t)$, yielding the SDE
\begin{equation}\label{eq:normalization_SDE}
 d \mathcal{Q} = \mathcal{Q} \cdot \left( \frac{2}{\hbar} \Im\langle \mathcal{E}_\text{eff}^\text{lin} \rangle dt + 2 \sum_n \Re \langle \mathcal{B}^{n} \rangle \circ dW_n^{} \right) \,\text{.}
\end{equation}
The dynamics of the stochastic variable $\mathcal{Q}$ can be integrated along with the dynamics of the set of parameters $\boldsymbol{\theta}$. Then, the expectation value of an observable $\hat A$ at a certain time is calculated by the following average over $N_\text{T}$ quantum trajectories:
\begin{equation}\label{eq:expectation_linear}
 \llangle \hat A \rrangle = \frac{1}{N_\text{T}} \sum_j \frac{ \langle \psi_{\boldsymbol{\theta}_j} | \hat A | \psi_{\boldsymbol{\theta}_j} \rangle } { \langle \psi_{\boldsymbol{\theta}_j} | \psi_{\boldsymbol{\theta}_j} \rangle } \cdot \mathcal{Q}_j \,\text{,}
\end{equation}
where $\boldsymbol{\theta}_j$ is the set of variational parameters of the $j^\text{th}$ trajectory, and $\mathcal{Q}_j$ is the corresponding normalization. The normalized expectation value $\langle \psi_{\boldsymbol{\theta}_j} | \hat A | \psi_{\boldsymbol{\theta}_j} \rangle / \langle \psi_{\boldsymbol{\theta}_j} | \psi_{\boldsymbol{\theta}_j} \rangle$ is estimated by a Monte Carlo average, while $\mathcal{Q}_j$ is obtained by numerically integrating the SDE~\eqref{eq:normalization_SDE}.

We find that the expectation values obtained by solving the otVMC equations for the linear SSE exhibit significantly higher noise compared to those from the nonlinear SSE. This is likely due to the additional integration required for the auxiliary variable $\mathcal{Q}$, as well as the use of a weighted average in Eq.~\eqref{eq:expectation_linear}, which introduces further statistical fluctuations.

\section{SDE solvers}\label{app:solver}

Finite difference solvers for SDEs of form $dX = A dt + B \circ dW_t$ can be derived from a stochastic Taylor expansion to the desired order in the time step $\Delta t$, taking into account that $\Delta W_t$ is $O(\Delta t^{1/2})$. 
However, this would introduce terms of type $\partial B/\partial X$, which are cumbersome to evaluate with VMC. Schemes more suited to our approach are of predictor-corrector type. A predictor step of Euler-Maruyama type $\bar{X}=X_i + A(X_i)\Delta t +B(X_i)\Delta W_i$ is performed, where subscript $i$ is the step index. Then, a corrector step provides an accurate estimation of $X_{i+1}$ employing quantities evaluated at the predictor step. We tested two approaches: in the trapezoidal-rule method~\cite{Kloeden_NumericalSolutionStochastic_1995}, one evaluates $X_{i+1}=X_i + (1/2)(A(X_i)+A(\bar{X}))\Delta t + (1/2)(B(X_i)+B(\bar{X}))\Delta W_i$; in the midpoint-rule method~\cite{Milstein_NumericalMethodsStochastic_2002}, one evaluates $X_{i+1}=X_i + A((X_i+\bar{X})/2)\Delta t + B((X_i+\bar{X})/2)\Delta W_i$. These schemes are both compatible with the Stratonovich interpretation, differently than a single-step Euler-Maruyama scheme, and are of strong convergence order 1, in the case of a single noise source. In our more general multiple noise case, they are both of strong convergence order $1/2$~\cite{Kloeden_NumericalSolutionStochastic_1995}: we verified that both have comparable accuracy and cost, and, in this work, all shown results are obtained using the midpoint method.

\section{Exact results for the 1D LITF model}\label{app:fossfeig}
For the LITF without transverse field, we employ the analytical solution from \cite{FossFeig_Nonequilibriumdynamicsarbitraryrange_2013} and take advantage of the translational invariance of the coupling, defining $J_d=(J/\mathcal{K}(\alpha)) |d|^{-\alpha}$ for $0<d<N/2$, $J_d=J_{N-d}$ for $d\ge N/2$ and $J_0=0$. We consider a quench starting from the coherent spin state along $+x$, set $\Gamma=\kappa/2$, and define the following auxiliary quantities:
\begin{align}
 \Phi(a) &= e^{-\Gamma t}\left(\cos{\left(2a t+i \Gamma t\right)}+\frac{\Gamma}{2a + i \Gamma}\sin{\left(2a t+i \Gamma t\right)}\right)\nonumber\\
 \Psi(a) &= 2e^{-\Gamma t}\frac{-\Gamma+i a}{2a + i \Gamma}\sin{\left(2a t+i \Gamma t\right)}\nonumber\\
 \Phi^\mu &= \prod_{d=0}^{N-1}\Phi(\mu J_d)\quad \text{for}\;\mu=\pm 1 \nonumber \\
 \Phi^{\mu\nu}_\Delta &= \prod_{d=0}^{N-1}\Phi(\mu J_d+\nu J_{\Delta-d})\quad \text{for}\;\mu,\nu=\pm 1 \nonumber \\
 Q_{\mu z} &= \frac{e^{-\Gamma t}}{2N} \Phi^\mu \sum_{d=1}^{N-1}{\frac{\Psi(\mu J_d)}{\Phi(\mu J_d)}} \nonumber \\
 Q_{\mu\nu} &= \frac{e^{-2\Gamma t}}{4N} \sum_{\Delta=1}^{N-1}{\frac{\Phi^{\mu\nu}_\Delta}{\Phi(\mu J_\Delta)\Phi(\nu J_\Delta)}}\;.
\end{align}
We thus obtain the solution for the average collective magnetization $(M_x,\;M_y,\;M_z) = (e^{-\Gamma t}\Re \Phi^+,\; e^{-\Gamma t}\Im \Phi^+,\; e^{-2\Gamma t}-1)$, and for the normalized spin covariance:
\begin{align}
 C_{xx} &= 1/N + \Re{( 2Q_{++}+Q_{-+}+Q_{+-})} - M_x^2 \nonumber\\
 C_{yy} &= 1/N + \Re{( -2Q_{++}+Q_{-+}+Q_{+-})}-M_y^2 \nonumber\\
 C_{xy} &= 2\Im{Q_{++}}-M_x M_y \nonumber\\
 C_{xz} &= \Re{(Q_{+z}+Q_{-z})}-M_x M_z \nonumber\\
 C_{yz} &= \Im{(Q_{+z}-Q_{-z})} -M_y M_z \nonumber\\
 C_{zz} &= (1/N)\left(1-M_z^2 \right)\;.
\end{align}

\section{Computational cost}\label{app:cost}
\begin{table}[hbp]
\centering
\begin{tabular}{r|c|c|c|c|c}
 N & LJH & LJPH1 & LJPH2 & LJNH & RBM \\ 
 \hline
 10 & 7\text{s} & 9\text{s} & 13\text{s} & 25\text{s} & 2\text{min} \\
 100 && 15\text{min} &&& \\
 160 && 30\text{min} &&& 
\end{tabular}
\caption{Wall time duration of typical trajectories for total time $T_{\text{max}}\approx 0.8/J$. See text.}\label{tab:cost}
\end{table}
In our method, both tVMC sampling and the quantum trajectories can be parallelized, and our code currently employs MPI on multiple CPUs. We collect in Table~\ref{tab:cost} the typical wall times for single quantum trajectories simulated in this work, with tVMC sampling distributed on 72 cores @2.5GHz. The LJPH1 ansatz simulations scale approximately as $N^2$.
\newpage
\end{document}